\documentclass{article}

\usepackage{arxiv}

\usepackage[utf8]{inputenc} % allow utf-8 input
\usepackage[T1]{fontenc}    % use 8-bit T1 fonts
\usepackage[bookmarks=false]{hyperref}       % hyperlinks
\usepackage{url}            % simple URL typesetting
\usepackage{booktabs}       % professional-quality tables
\usepackage{amsfonts}       % blackboard math symbols
\usepackage{nicefrac}       % compact symbols for 1/2, etc.
\usepackage{microtype}      % microtypography
\usepackage{lipsum}		% Can be removed after putting your text content
\usepackage[authoryear,square,comma]{natbib}
\usepackage{doi}

\usepackage[T1]{fontenc}
\usepackage[pdftex]{graphicx} 
\usepackage{siunitx}

\usepackage{amsmath,mathtools}
\usepackage{amsfonts}
\usepackage{algorithm}
\usepackage[noend]{algpseudocode}
\usepackage{eqparbox}
\usepackage{accents}
\newcommand{\ubar}[1]{\underaccent{\bar}{#1}}
\newcommand{\determ}[1]{\mathrm{det}(#1)} % determinant

\newdimen{\algindent}
\algnewcommand\LeftComment[2]{%
\hspace{#1\algindent}$\triangleright$ \eqparbox{COMMENT}{#2} \hfill %
}

\usepackage{graphicx,amsmath}

\newcommand{\vect}[1]{\mathbf{#1}}
\DeclareMathSymbol{\shortminus}{\mathbin}{AMSa}{"39}

\title{ Sensor Placement for Flapping Wing Model\\ Using Stochastic Observability Gramians}

\author{
	\href{https://orcid.org/0000-0002-3268-1467}{\includegraphics[scale=0.06]{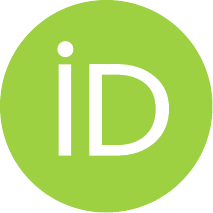}\hspace{1mm}Burak ~Boyacıoğlu} \\
	University of Washington \\
	Seattle, WA 98195 \\
	\texttt{burakb@uw.edu} \\
	\And
 \href{https://orcid.org/0000-0001-5097-5672}{\includegraphics[scale=0.06]{orcid.pdf}\hspace{1mm}Mahnoush ~Babaei}\\
	University of Texas at Austin \\
	Austin, TX 78712 \\
	\texttt{mahnoushb@utexas.edu} \\
	\And
	Amanuel ~H. ~Mamo \\
	University of Washington \\
	Seattle, WA 98195 \\
	\texttt{amanuel9@uw.edu} \\
	\And
	\href{https://orcid.org/0000-0003-2735-0206}{\includegraphics[scale=0.06]{orcid.pdf}\hspace{1mm}Sarah ~Bergbreiter} \\
	Carnegie Mellon University \\
	Pittsburgh, PA 15213 \\
	\texttt{sbergbre@andrew.cmu.edu} \\
	\And
	\href{https://orcid.org/0000-0002-5706-1096}{\includegraphics[scale=0.06]{orcid.pdf}\hspace{1mm}Thomas ~L. ~Daniel} \\
	University of Washington \\
	Seattle, WA 98195\\
	\texttt{danielt@uw.edu} \\
	\And
	\href{https://orcid.org/0000-0001-9826-7886}{\includegraphics[scale=0.06]{orcid.pdf}\hspace{1mm}Kristi ~A. ~Morgansen} \\
	University of Washington \\
	Seattle, WA 98195 \\
	\texttt{morgansn@uw.edu} \\
}

\renewcommand{\shorttitle}{Sensor Placement Using Stochastic Gramians}

\hypersetup{
pdftitle={Sensor Placement for Flapping Wing Model Using Stochastic Observability Gramians (Preprint)},
pdfauthor={Burak~Boyacioglu, Mahnoush~Babaei, Amanuel~H.~Mamo, Sarah~Bergbreiter, Thomas~L.~Daniel, and Kristi~A.~Morgansen},
}

\begin{document}
\maketitle

\begin{abstract}
Systems in nature are stochastic as well as nonlinear. In traditional applications, engineered filters aim to minimize the stochastic effects caused by process and measurement noise. Conversely, a previous study showed that the process noise can reveal the observability of a system that was initially categorized as unobservable when deterministic tools were used. In this paper, we develop a stochastic framework to explore observability analysis and sensor placement. This framework allows for direct studies of the effects of stochasticity on optimal sensor placement and selection to improve filter error covariance. Numerical results are presented for sensor selection that optimizes stochastic empirical observability in a bioinspired setting.
\end{abstract}

% keywords can be removed
\keywords{ Observers for nonlinear systems \and Computational Methods  \and Stochastic Systems}

\section{Introduction}
Performance resilience in the presence of sensing noise, environmental disturbances, and modeling imprecision is typically approached from the perspective of robust control theory. 
However, all feedback-based control techniques, including robust methods, are constrained by the quality of the state estimate as determined both by the accuracy of the estimate (how close the estimate is to the true value) and the precision (how well the estimate is known).  Specifically, the precision, as captured by the covariance of the error, directly affects the size of the robustness margin of the controller.  Even in linear systems where the separation principle indicates that the design of system control and estimation are independent, the structural properties of sensor location have a direct impact on control performance.  As shown in \cite{Powel2015}, observability in linear systems directly bounds the error covariance for estimation filters.  Improving the system observability necessarily improves the error covariance, regardless of the particular filter being used.  These improvements can be further enhanced by assessing observability in the presence of noise using stochastic tools rather than the typical deterministic observability framework. The work in this paper addresses the use of stochastic observability tools in the task of determining optimal sensor placements, particularly in the context of sparse sensing.

System observability tools determine whether or not state values can be uniquely obtained via finite-time measurements. The observability of nonlinear systems can be affected by the control inputs and should be examined considering the input information. However, in most nonlinear systems of interest, analytical calculation of observability is impractical. In such cases, the empirical Gramians, which were originally introduced for model reduction \cite{Lall1999}, provide a good sense of controllability/observability as their eigendecompostions define controllable/observable spaces. Inspired by measures based on the linear Gramians \cite{muller1972} and the observability matrix \cite{Dochain1997}, empirical Gramian-based observability measures have been studied to optimize fixed-sensor placement \cite{Qi2014} or vehicle trajectory planning \cite{hinson2013b,Glotzbach2014} for data collection. Recently, an empirical Gramian rank condition for weak observability of time-invariant nonlinear systems was developed \cite{Powel2015}, and a powerful open-source toolbox for the calculation of Gramians has been introduced for both time-varying and time-invariant systems \cite{himpe2018}.

The empirical Gramian has also been studied for stochastic systems. In \cite{powelArXiv}, it was shown that if an unobservable system is simulated with process noise, one may get nonzero observability Gramian eigenvalues, that is, considering uncertainty reveals the observability of the system which is initially determined unobservable using traditional methods. Although a rank condition for the expected value of the empirical observability Gramian for stochastic observability was given for linear systems in the same study using the stochastic observability definition from \cite{Dragan2004}, it was stated that an extension to nonlinear systems was not readily available since the fundamental matrix of linear systems has no analog in the nonlinear context.

In this paper, we study the optimal sensor placement problem using stochastic observability measures and show the cost function should be chosen according to the system structure. We consider two example applications. The first is a low-dimensional example of an unmanned aerial vehicle (UAV) wind tracking problem, and the second is a flapping wing model. Because dynamic models based on Euler-Lagrange equations take only a limited number of modes of the structure into account \cite{eberle2015,hinson2015},
we have created a finite element analysis (FEA) model of the wing. The output model we use is inspired by the neural encoding mechanism in animal sensing.

The remainder of the paper has been organized as follows. Section~\ref{sec:back} gives the background of observability analysis tools and  observability measures. Section \ref{sec:model1} and \ref{sec:model2} summarize the example system models.
%wing model with the adopted neural encoder.
Section~\ref{sec:method} describes the optimal sensor placement methodology.
Simulation results %using the FEA model 
are included in Section~\ref{sec:sim}.
Finally, Section~\ref{sec:last} gives the conclusions and directions for future work.

\section{BACKGROUND}\label{sec:back}

Here, we summarize the relevant materials in analytical and empirical observability.

\subsection{Observability Analysis}
 Consider the continuous-time linear control system,
  \begin{equation}\label{wo_memory}
  \begin{split}
   {\vect{\dot x}}(t) &= A\vect{x}(t)+B\vect{u}(t)\\
   \vect{y}(t) &= C\vect{x}(t),\\
  \end{split}
 \end{equation}
where $\vect{x}\in\mathbb{R}^n$ is the state vector, $\vect{u}\in\mathbb{R}^m$ is the input vector, and $\vect{y}\in\mathbb{R}^p$ is the output vector. The existence of a unique mapping from the output space over a finite time interval to the state space is called observability of a system, and this property can be tested for the system (\ref{wo_memory}) by checking the rank of the observability matrix,
\begin{equation}\label{observabilityk}
    \mathcal{O}_T=\begin{bmatrix}C\\ CA\\\vdots\\ CA^{T-1} \end{bmatrix}
\end{equation}
for $T\geq n$. If $\mathcal{O}_T$ has full column rank, then the system is observable. Otherwise, the desired unique mapping does not exist, and the system is unobservable \cite{sontag1990}.

An equivalent rank condition can be determined using the observability Gramian,
\begin{equation}\label{linGram}
    W_o=\int_{0}^\infty (e^{\tau A^\top} )^tC^\top Ce^{\tau A}d\tau \in \mathbb{R}^{n \times n}.
\end{equation}
Here, $\top$ denotes the matrix transpose. For observability, $W_o$ must have rank $n$. Notice that neither of the conditions is affected by the input term. This separation principle usually does not hold for nonlinear systems. Also note that the observability Gramian is at least semi-definite by construction; hence, the eigenvalues of $W_o$ are equal to its singular values.

\subsection{Deterministic Nonlinear Observability}

The fundamentals of  observability analysis of nonlinear systems in  control-affine form is based on  differential geometric techniques \cite{Anguelova2004}. However, the analytical calculations quickly become intractable for most physical systems of interest. A numerical tool termed the empirical Gramian provides more flexibility for  nonlinear observability analysis by considering a local assessment of observability from a local linearization about a nominal trajectory.  This approach benefits from not requiring analyticity of the dynamics, but does so at the expense of losing strict guarantees.

Consider the nonlinear system with process noise,
 \begin{equation}\label{nonlinear}
   \begin{split}
     \dot{\vect{x}}(t) &= \vect{f}(\vect{x}(t),\vect{u}(t))+G\mathbf{w}(t)\\ 
     \vect{y}(t) &= \vect{h}(\vect{x}(t)),\\
  \end{split}
 \end{equation}
where $\mathbf{w}$ is the vector of noise components, and the full dynamic model may be partially unknown, as in the case of an FEA model or a data-based model with high dimensions. To obtain the deterministic empirical Gramian with $\mathbf{w}=\mathbf{0}$, the initial condition of each state variable is perturbed independently by amount $\mp\varepsilon$ and the system is simulated with an input sequence, $\vect{u}\in\mathcal{U}$, where $\mathcal{U}$ is the set of permissible controls. Let
\begin{equation}
    \vect{y}^{\mp i}(t)=\vect{h}(\vect{x}(t,\vect{x}_0\mp \varepsilon \vect{e}_i,\vect{u})),
\end{equation}
where $\vect{e}_i$ denotes the $i$-th standard basis vector in $\mathbb{R}^n$. Then the deterministic empirical observability Gramian is obtained as
\begin{equation}
\label{eq:gramian}
W_o^\varepsilon(t_1, \mathbf{x}_0, \vect{u}) = \frac{1}{4\varepsilon^2} \int_0^{t_1} \Phi^\varepsilon(t,\mathbf{x}_0,\vect{u})^\top \Phi^\varepsilon(t,\mathbf{x}_0,\vect{u}) dt,
\end{equation}
where $t_1$ is the simulation time, $\mathbf{x}_0$ is the initial state and
\begin{equation}
\label{eq:phi}
\Phi^\varepsilon(t,\mathbf{x}_0,u) = \begin{bmatrix}\mathbf{y}^{+1}-\mathbf{y}^{-1} & \cdots & \mathbf{y}^{+n}-\mathbf{y}^{-n}\end{bmatrix}.
\end{equation}

A rank condition for the empirical Gramian analogous to the condition given for the linear Gramian was introduced in \cite{Powel2015}. However, since it requires obtaining the limit value of (\ref{eq:gramian}) as $\varepsilon \to 0$, it is usually not possible to determine the observability precisely. In that case, an approximate lower bound on the minimum singular value of the matrix can be calculated \cite{powelArXiv}.

\subsection{Observability Measures}
Krener and Ide \cite{Krener2009} introduced two unobservability measures based on the empirical Gramian: the reciprocal of the minimum eigenvalue of the observability Gramian, $1/\ubar{\lambda}(W_o)$, which is also called the unobservability index, $\nu(W_o)$, and the condition number of the same matrix, $\kappa(W_o)= \bar{\lambda}(W_o)/\ubar{\lambda}(W_o)$. The former points out the least observable direction similar to determining the weakness of a chain by its least strong link. The latter shows the balanced contribution of state variables to the output, and its value is desired to be one assuming that the output coordinates are already scaled. Although, as pointed out in \cite{Krener2009}, $\mu=1/\kappa(W_o)$ does not necessarily increase as new information added to the system, the condition number of the observability Gramian is still a significant measure as it shows how well-conditioned the estimation problem is.

In \cite{Qi2014}, the optimal phasor measurement unit (PMU) placement problem was formulated to maximize the determinant of the observability Gramian, $\determ{W_o}$, which is equal to the product of all the eigenvalues of $W_o$, but it was advised to check the minimum eigenvalue to be at an acceptable level. % In \cite{Boyacioglu2021}, we calculated an approximate value of such a level for the hawkmoth flapping wing dynamics without noise using the theory and numerical method presented in \cite{Powel2015,powelArXiv}.
Since the maximization of $\determ{W_o}$ is not a convex problem, $\log\determ{W_o}$ is sometimes preferred instead, e.g., in \cite{Serpas2013}. Here, we adopt the $n$-th root of the determinant, $[\determ{W_o}]^{1/n}$, which is not only concave but, as stated in \cite{Singh2005}, also equals zero when the system is unobservable and yields a positive number when it is not.

%\bb{\sout{Although the trace and the spectral radius might be useful for applications like model reduction, they are not usually the first choice for observability analysis as they are not able to catch the case when $W_o$ is singular.}}

\subsection{Stochastic Empirical Gramian}
In order to address the assessment of observability in the presence of noise, the stochastic empirical Gramian was introduced in \cite{powelArXiv} with two calculation methods. One method is computing new sample trajectories for each entry of the Gramian; in this case the total number of simulations is $4n^2$, and the Gramian is not guaranteed to be at least positive semi-definite. In this study, we adopt the second method and run $2n$ simulations in total, assuming that $\Phi^\varepsilon(t,\mathbf{x}_0,\mathbf{u},\mathbf{w}^{+1},\mathbf{w}^{-1},\dots,\mathbf{w}^{+n},\mathbf{w}^{-n})$ and its transpose are obtained from the same simulations. Here, $\mathbf{w}^{\mp i}$ is the noise sequence used to obtain the output $\mathbf{y}^{\mp i}$, i.e.,
\begin{equation}
    \vect{y}^{\mp i}(t)=\vect{h}(\vect{x}(t,\vect{x}_0\mp \varepsilon \vect{e}_i,\vect{u}),\vect{w}^{\mp i}).
\end{equation}

Unlike the empirical Gramian for deterministic systems, the spectrum, or eigenvalues, of the stochastic Gramian would vary with each build, even though the initial state, the input sequence, and the simulation time remain the same. Additionally, as illustrated in \cite{powelArXiv} for two different systems, measures such as the condition number and the minimum eigenvalue cannot be assumed to follow Gaussian distributions when the system is simulated numerous times in the presence of Gaussian process noise. Consequently, the expected value of a Gramian-based observability measure usually cannot be obtained analytically due to the complex dynamics.

In \cite{powelArXiv}, it was also illustrated that the unobservability index decreases as the noise variance increases which  is intuitive as the output energy will be higher although there is more uncertainty. On the other hand, the estimation condition number is not a monotonic function of the noise level. Finally, we simulated both systems from \cite{powelArXiv} and observed that the reciprocal of the $n$-th root of the determinant of the empirical Gramian has similar behaviour to the unobservability index.

%\newpage
\section{UAV NAVIGATION MODEL}\label{sec:model1}
We will explore the work here relative to two examples, the first being low-dimension dynamics of a UAV system, and the other a flapping wing system with flexible structure described by infinite dimensional modal dynamics.

%In order to investigate the choice of unobservability measure, $j$, with a large number of simulations, we chose the unmanned aerial vehicle (UAV) navigation problem. 
In \cite{hinson2013b}, a simplified planar model of a nonholonomic fixed-wing aircraft in the presence of air currents is given as:
\begin{equation}\label{sys_uav}
\dot{\mathbf{x}}=\left[\begin{array}{c}{V \cos x_{3}+x_{4}} \\ {V \sin x_{3}+x_{5}} \\ {u} \\ {0} \\ {0}\end{array}\right]+\left[\begin{array}{ll}1&{0} \\ {0}&1 \\ {0}&0 \\ {0}&0 \\ {0}&0\end{array}\right] \mathbf{w}, \quad \mathbf{y}=\left[\begin{array}{l}{x_{1}} \\ {x_{2}}\end{array}\right],
\end{equation}
where $\mathbf{x}=\left[\begin{array}{lllll}{x_{E}} & {y_{N}} & {\theta} & {W_{x}} & {W_{y}}\end{array}\right]^{\top}$. The aim is to estimate the wind components in $x$ and $y$ directions, $W_x$ and $W_y$. Here, $x_E$ and $y_N$ are the vehicle's inertial East and North positions, and $\theta$ is its inertial orientation. $V$ is the vehicle flow-relative velocity, which is assumed to be constant, and the control input, $u$, is the vehicle's angular velocity. Finally, $\mathbf{w}$ is the vector of independent and identically distributed (i.i.d.) zero-mean Gaussian process noises with  diagonal  covariance matrix, $Q \in \mathbb{R}^{2 \times 2}$.

\section{FLAPPING WING MODEL}\label{sec:model2}
%\nbnote{no intro paragraph for this section?}

In order to explore the determination of optimal sensor placement in a higher dimension stochastic framework using empirical Gramian methods, we will frame the work around a biologically inspired flapping flight example. The dynamics of a continuum flexible system represented in modal coordinates, $\eta$, can be expressed as
\begin{equation}
M \ddot{\boldsymbol{\eta}} + C\dot{\boldsymbol\eta} + K \boldsymbol\eta= \mathbf{u}(t),%\quad \mathbf{y}=\mathbf{H}(\vect{\eta})    ,
\end{equation}
where $\boldsymbol\eta \in \mathbb{R}^n$ are the structural modes, $M$ is the mass matrix, $C$ is the damping matrix, $K$ is the stiffness matrix, and $\mathbf{u}$ are the external forces.  We define the augmented system state vector as the mode shapes, their time derivatives and the flapping rate, $\dot\phi$ and rotation rate, $\omega$, of the wing:
\[
{\mathbf x} = \begin{bmatrix} \dot{\phi} & \omega & \boldsymbol\eta^\top  & \dot{\boldsymbol\eta}^\top 
\end{bmatrix}^\top .
\]  The overall dynamics can then be expressed as
\begin{equation}
    \dot{\mathbf{x}} = \mathbf{f}({\mathbf x}, {\mathbf u}(t))  +\begin{bmatrix} 1&0\\0&1\\0 &0\\\vdots&\vdots\end{bmatrix} \mathbf{w},\quad \mathbf{y}=\mathbf{h}(\vect{\bf{x}}) ,
\end{equation}
where $\mathbf{w}\sim\mathcal{N}(\vect{0},Q)$. The system outputs are assumed to be provided from a neural encoding model described below.

\subsection{FEA Model}
\begin{figure}[tb!]
\centering
\includegraphics[width=0.7\linewidth]{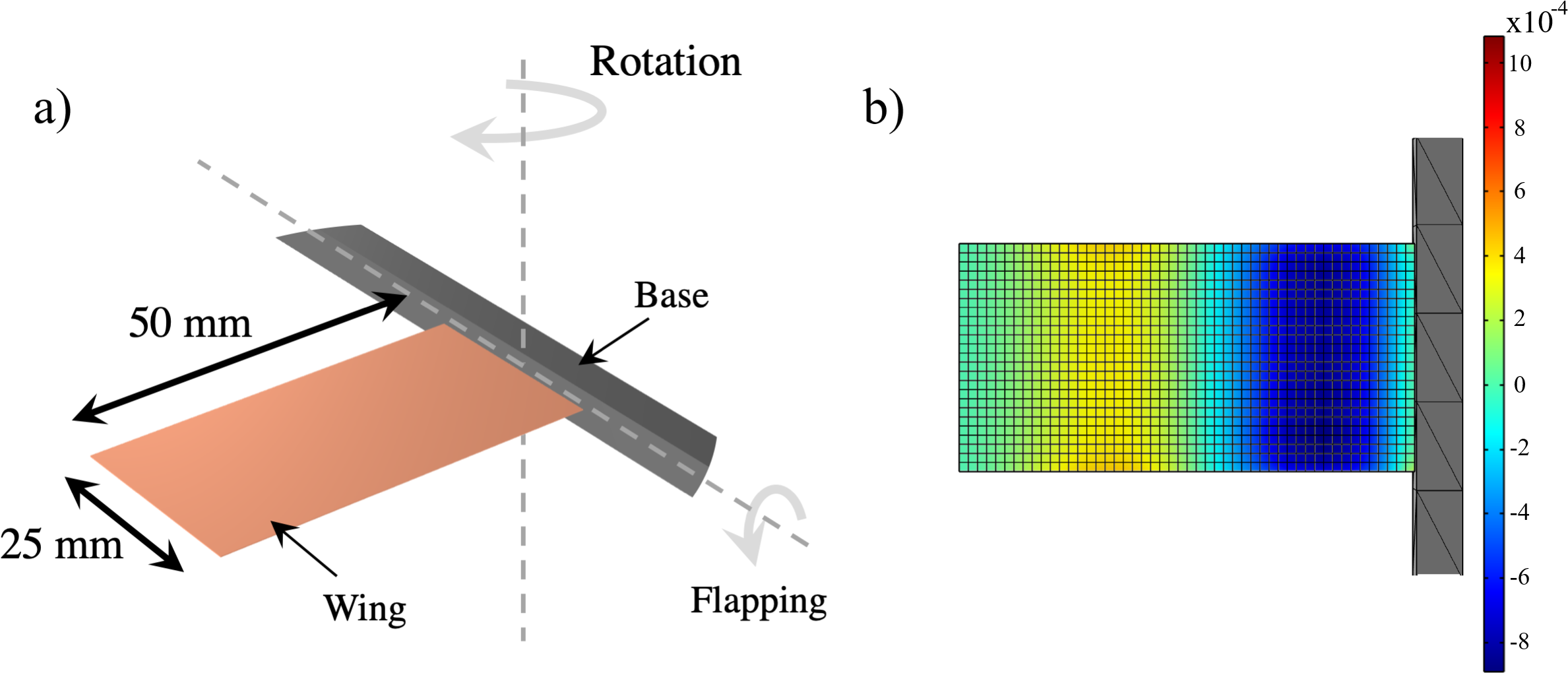}
\caption{(a) Details of structural model and boundary conditions for flapping wing model in COMSOL, (b) meshing of the wing plate and normal strain distribution on the top surface of wing.}
\label{FEAmodel}
\end{figure}

To study the dynamics and the observability of a flapping wing, we created an FEA model of the wing in COMSOL Multiphysics\textsuperscript{\footnotesize\textregistered} 5.6. We generated this simplified model based on the properties of the hawkmoth \textit{Manduca Sexta} wing. As shown in Fig. \ref{FEAmodel}a, the structure was modeled as a single rectangular thin plate with a width of \SI{25}{\milli\metre}, length of \SI{50}{\milli\metre}, and thickness of \SI{12.7}{\micro\metre}. We considered the thickness of the structure to be uniform throughout the wing, and the effect of venation was neglected in the simulations. To define the Young’s modulus ($E$) of the wing, we chose a value within the range of stiffness values measured previously for insect wings \cite{Combes2979}. Values of the material properties used in the FEA model are shown in Table \ref{tab_sim}. 

The damping has a significant effect on the dynamic response and the strain patterns generated due to flapping. Hence, we adopted a Rayleigh damping model with damping matrix defined as a linear combination of the mass and the stiffness matrices specified using coefficients $\alpha$ and $\beta$:
\begin{equation}
\label{eq:damp}
   C=\alpha M+\beta K.
\end{equation}
The wing was modeled as linearly elastic and was meshed using 25 by 50 structured quadrilateral elements swept through the thickness of the plate (Fig. \ref{FEAmodel}b). The mesh size was selected based on a mesh convergence analysis to ensure the accuracy of the results and minimum computation time.

We applied boundary conditions to mimic the flapping pattern of \textit{Manduca} using the steady flapping model from \cite{Mohren10564}. The model has an amplitude of $\pi/6$ radians and consists of two harmonic components to match the experimental results. The first component acts at a frequency of $f_1=$\SI{25}{\hertz}, while the second component with $1/5$ of the amplitude has a frequency of $f_2=$\SI{50}{\hertz}. We introduced flapping in terms of angular velocity, that is,
\begin{equation}
\label{eq:flap}
\dot{\phi}(t)=\frac{\pi}{6}(2\pi f_1cos(2\pi f_1t)+\frac{2\pi f_2}{5}cos(2\pi f_2t)).
\end{equation}

This equation is obtained by taking the time derivative of the function describing steady flapping angle at the base of the wing, as described above. 
In addition to flapping, we introduced a perturbation in the form of inertial rotation as shown in Fig.\ref{FEAmodel}a. Although the nominal trajectory involves no rotation, we were required to add a rotation of small magnitude ($\omega=\SI{0.02}{\radian\per\second}$) to ensure convergence of the model. This rotation will also impose asymmetry, which naturally comes from the asymmetrical
shape and stiffness of the wing.

The random function in COMSOL software allowed us to use a random seed that is guaranteed to generate a new noise sequence while maintaining the same distribution. Proportionally to the nominal trajectory, we picked $Q=\mathrm{diag}{(1,1\times10^{-4})}(\SI{}{\radian\per\second})^2$ where the larger variance corresponds to the noise on the flapping rate.

\begin{table}[t]
    \centering
    \caption{Material Properties Used in COMSOL Simulations}
    \begin{tabular}{c|c}
        \textbf{Properties} & \textbf{Values}  \\
                % \vspace{5 pt}
        \hline
        Young's Modulus ($E$) & \SI{0.3}{\giga\pascal} \\
        Poisson's Ratio & 0.35 \\
        Mass Damping Coefficient ($\alpha$) & \SI{500}{\per\second} \\
        Stiffness Damping Coefficient ($\beta$) & \SI{0}{\second} \\
        Density & \SI{1180}{\kilo\gram\per\m^3}
    \end{tabular}
    \label{tab_sim}
    \vspace{-10pt}
\end{table}

%\textcolor{teal}{In the deterministic case, we can immediately tell that the rotation direction is not observable on the x-axis since the current wing model is symmetrical about it, which will not be a concern anymore once the symmetry is removed, e.g. when the profile of the insect wing is used.}
For each simulation, we extracted the normal strain along the length of the wing (Fig. \ref{FEAmodel}b) to feed into the neural encoding model, which is outlined in the following subsection.

\subsection{Neural Encoding Measurement Model}
We used a neural encoding model \cite{pratt2017} to obtain the probability of firing of a neuron from the strain information at the location of the corresponding sensor. The model consists of a linear filter, similar in concept to moving average filters used in engineered systems, and a nonlinear activation function, NLA. The filter coefficients are determined using the spike-triggered average (STA),
\begin{equation}\label{stave}
    \operatorname{STA}(t)=\cos{\left(2\pi f_\text{STA}(-t+a)\right)}\operatorname{exp}{\left(\frac{-(-t+a)^2}{b^2}\right)}.
\end{equation}
Here, $b$, $f_\text{STA}$, and $a$ are the width, the STA frequency, and the delay, respectively. The similarity metric for a new stimulus, $\epsilon$, with the STA is obtained by convolving them and scaling by a constant, $C_\xi$:
\begin{align}\label{strain_proj}
    \xi(x,y,t)=\frac{1}{C_\xi}\int_{0}^{N}\epsilon(x,y,t-\tau)\cdot STA(\tau) d\tau.
\end{align}
The NLA function determines the probability of firing, $P_{\text{fire}}$, based the projected stimulus, $\xi$:
\begin{equation}\label{probFire}
    P_{\text{fire}} =NLA(\xi)=\frac{1}{1+\exp (-c(\xi-d))}.
\end{equation}
Here, $c$ is the slope and $d$ is the half-maximum position of the $\operatorname{NLA}$ function; the $x$, $y$ and $t$ arguments have been dropped for brevity. Notice that $P_{\text{fire}}$ at a time $t$ is not only a function of the strain value at that time but also includes an output delay with a length $N$.

\begin{figure}[t!]
\centering
\includegraphics[width=0.6\linewidth]{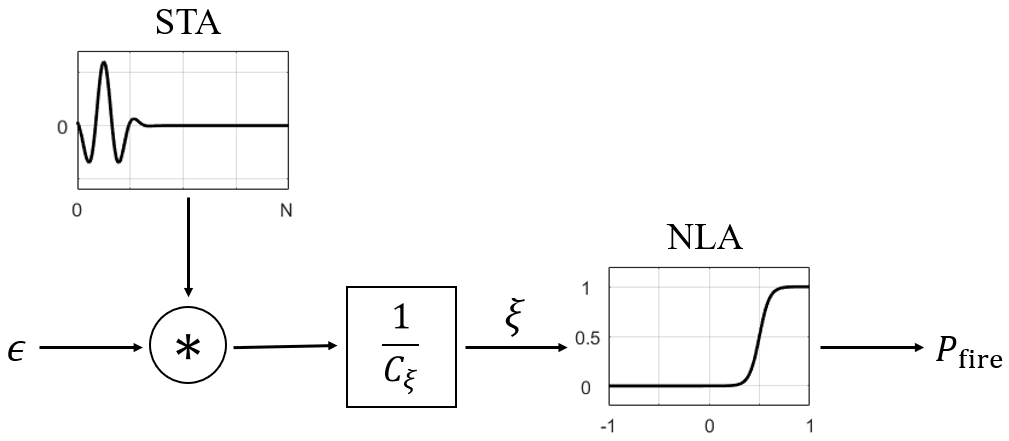}
\caption{The probabilistic firing model of a neuron where the input is the strain information, $\epsilon$, and the output is the probability of firing, $P_\text{fire}$ \cite{Boyacioglu2021}.}\label{NEmodel}
\end{figure}

The full neural encoding process is summarized in Fig. \ref{NEmodel}. In our application, the input to the neural encoding model is the strain at a location on the wing, and the output is $P_{\text{fire}}$. The encoder is experimentally validated in \cite{pratt2017}, and the parameters are regarded as constant for the neuron's model. By considering $P_\text{fire}$ as the output of the system, we keep the neural encoding model deterministic. This way, the only source of uncertainty will be the process noise.

\section{SENSOR PLACEMENT METHODOLOGY}\label{sec:method}
Based on the output energy from different sensor locations of a system, one could expect different levels of observability. In this section, we present a sensor placement methodology based on the spectrum of the stochastic observability Gramian.
To obtain the information of interest, we perform Monte Carlo analysis where the number of simulations was chosen based on convergence of the results.

Let $j:\mathbb{R}^{n\times n}\to \mathbb{R}$ be an unobservability measure as a function of the observability Gramian. Then the mean value of this measure obtained from $K$ simulations,
\begin{equation}
    \bar J(t_1,\mathbf{x}_0,\vect{u},\mathbf{w}^{+1}_1,\dots,\mathbf{w}^{-n}_1,\dots,\mathbf{w}^{+1}_K,\dots,\mathbf{w}^{-n}_K):=\frac{1}{K}\sum_{i=1}^Kj((W_o^\varepsilon)_i),
\end{equation}
should be minimized for the sensor placement problem. In other words, if $p$ is the finite number of possible sensor locations and we define ${(W_o^\varepsilon)_{k,i}}$ as the Gramian matrix obtained at the $k$-th sensor location on the $i$-th run, then
\begin{equation}(W_o^\varepsilon)_i(\boldsymbol{\gamma})=\sum_{k=1}^p\gamma_k{(W_o^\varepsilon)_{k,i}},
\end{equation}
where $\boldsymbol\gamma$ is the vector of Boolean-valued sensor activation functions, $\gamma_i$. The optimal sensor selection problem can be formulated as
\begin{equation}\label{eq:noncvxprob}
\begin{aligned}
& \min\limits_{\boldsymbol\gamma} 
& & \bar J(\boldsymbol\gamma) \\
& \text{subject to}
& &\sum\boldsymbol\gamma = r \\
& & &\gamma_k\in\{0,1\}, 
\end{aligned}
\end{equation}
where $r\leq p$ is the desired number of sensors. We are currently unaware of  a convex relaxation for $j$ if it is the condition number or the unobservability index when $K>1$.

To deal with the non-convexity of (\ref{eq:noncvxprob}), we applied the \textit{particle swarm optimization (PSO)} algorithm, a metaheuristic approach introduced in \cite{psopaper}. The PSO algorithm is a technique that generates several solution candidates, i.e., particles, and updates their position and velocity according to certain rules until a termination criterion is met \cite{psopaper}. Since this method is not problem-specific and allows abstract level description \cite{meta}, it is straightforward to implement. It also permits the definition of a search region, which is not the case for some widely-used algorithms such as the downhill simplex method \cite{neldermead}.

The pure PSO does not guarantee identification of the absolute local minimum. Hence, its hybridization with other optimization algorithms has been studied to find a better solution \cite{hybrids}. We also followed a hybrid approach, and at the expense of some computation time, used the result from PSO to initiate a second search by the \textit{interior point method (IPM)} discussed in \cite{interior}.

Note that the PSO algorithm requires a continuous search region, as opposed to be restricted to be able to have sensors at the $p$ potential locations in (14), and if each sensor has two coordinates to be determined, then the search space is $2r$-dimensional and the search would take more time as $r$ increases. To avoid having sensors too close to each other, we also implemented a relatively large positive value, $\sigma$, as the penalty for proximity.

The following pseudocode summarizes our approach to optimal sensor placement in a continuous space with upper and lower bounds, $ub$ and $lb$. Here, $d_{\min}$ is the minimum distance between any two of $r$ sensors, and $d_\text{allowed}$ is the minimum distance allowed. The PSO and IPM algorithm parameters have been suppressed for brevity. %\km{[what parameters do you mean?]}\bbnote{There are bunch of coefficients for PSO describing the certain rules we mentioned.}
\alglanguage{pseudocode}
\begin{algorithm}[h]
\caption{Optimal Sensor Placement}
\begin{algorithmic}[1]
\Procedure {}{dynamics, $ub$, $lb$, $j$, $K$, $r$, $d_\text{allowed}$, $\sigma$}\label{alg:pso}
\State \textbf{define} $(W_o^\varepsilon)_i \leftarrow \sum_{k=1}^r{(W_o^\varepsilon)_k}_i$
\State \textbf{define} $\hat J \leftarrow \begin{cases} 
      \frac{1}{K}\sum_{i=1}^Kj((W_o^\varepsilon)_i) & d_{\min}\geq  d_\text{allowed}\\
      \sigma & d_{\min}<d_\text{allowed}
   \end{cases}
$
\State sensor loci $\leftarrow$ PSO($\hat J$,$ub$, $lb$)
\State sensor loci $\leftarrow$ IPM($\hat J$,$ub$, $lb$, sensor loci)
\EndProcedure
\end{algorithmic}
\end{algorithm}
%\newpage
%\subsection{Application to UAV Navigation}

\section{NUMERICAL RESULTS}\label{sec:sim}

We now discuss the application results of the methods discussed here to the two example systems presented previously.  

\subsection{UAV Navigation System}
To build on the findings in \cite{powelArXiv} and further investigate the effects of noise, we study the system (\ref{sys_uav}) with no control and with five different levels of process noise. In prior work \cite{hinson2013b}, using a Lie algebraic approach for deterministic observability analysis, it is shown that this system is unobservable if $u(t)=\vect{0}$ and there is no process noise, i.e., $\vect{w}=\vect{0}$.

Since the first two states are directly measured, we select the remaining states as states of interest. For $K=100$ simulations with $\mathbf{x}_0=\left[\begin{array}{lllll}0& 0& \pi/6 & 0.35 & -0.15\end{array}\right]^{\top}$ and $t_1=\SI{150}{\second}$, allowing the nonzero process noise on $x_1$ and $x_2$ results in the stochastic empirical observability Gramian producing all nonsingular observability Gramians, indicating an observable system.

%$Q_{\text{ref}}=0.1I_2(\SI{}{\meter\per\second})^2$

\begin{figure}[hpbt!]
\centering \includegraphics[width=0.65\linewidth]{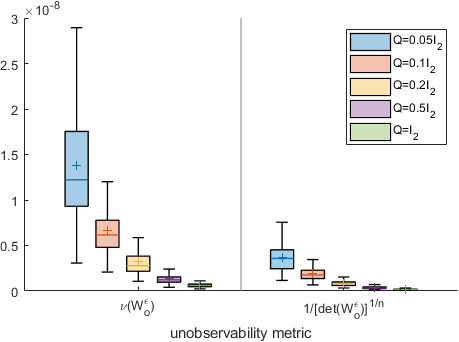}
\caption{The change of two unobservability metrics as the noise level increases. Here, horizontal lines with color indicate the median value, and the plus sign denotes the mean. $I_2$ is the $2 \times 2$ identity matrix.}\label{fig:uav_metric}
\end{figure}

Figure \ref{fig:uav_metric} shows the resultant unobservability measures $\nu(W_o^\epsilon)$ and $1/[\determ{W_o^\epsilon}]^{1/n}$. We observe a monotonic decrease in the unobservability index and its variance as the noise level increases. We also note the significant correlation between the two measures, $\rho=0.87$, although this correlation depends on the system dynamics itself as well as $Q$.%the covariance matrix.

%\begin{figure}[b]
%\centering
%\includegraphics[width=\linewidth]{figures/uavNav_5noiseLevels_I2_5e8.png}
%\caption{The cost of observability, $\hat J$, changing with the weight of the unobservability index for the five noise levels on the control-free UAV system.}\label{fig:uav_cost}
%\end{figure}

If we define the unobservability measure as a linear combination of the condition number and the unobservability index,
\begin{equation}
    j(W_o^\epsilon)=\kappa(W_o^\epsilon)+w_\nu \nu(W_o^\epsilon),
\end{equation}
then the weight of $\nu(W_o^\epsilon)$, $w_\nu$, will be of particular importance since $\kappa$ does not appear monotonic as a function of the noise, a characteristic which can be seen in Table \ref{tab:uav_cost} for $w_\nu=0$. Table \ref{tab:uav_cost} also illustrates this inference: as the unobservability index becomes dominant in the objective function, lower process noise on the system becomes less preferable. 
\begin{table}[ht]
    \centering
    \caption{Observability Cost of UAV Navigation}
    \begin{tabular}{@{}lcccccc@{}} % Remove vertical lines and adjust column spacing
        \toprule % Top horizontal line
        & \multicolumn{5}{c}{Process Noise Covariance, $Q$} \\ \cmidrule{2-6} % Partial horizontal line
        $w_\nu$ &$0.05I_2$ & $0.1I_2$& $0.2I_2$ & $0.5I_2$ & $I_2$ \\ \midrule % Mid horizontal line
        $0$ & $17.24$ & $16.34$ & $17.40$ & $15.69$ & $15.63$ \\
        $5\times10^8$ & $24.12$ & $19.66$ & $19.00$ & $16.36$ & $15.96$ \\ \bottomrule % Bottom horizontal line
    \end{tabular}\label{tab:uav_cost}
\end{table}

\subsection{Flapping Wing System}

For this work, we studied optimal neural-inspired sensor placement on the system described in Sec. \ref{sec:model2} aiming to increase observability.
Particularly inspired/informed by a previous study \cite{Mohren10564} that showed that the existence of externally induced body rotation as large as $\omega=\SI{10}{\radian\per\second}$ on a flapping wing differs by a twisting mode three orders of magnitude smaller than the dominant flapping mode.

We simulated the FEA model for \SI{0.2}{\second} (5 wingbeats) using time step of \SI{5e-4}{\second}. The motion was prescribed through angular velocities of flapping and rotation. Both velocities were zero through the first wingbeat cycle and linearly increased to their maximum values ($\sim$\SI{57}{\radian\per\second} for flapping and \SI{0.02}{\radian\per\second} for rotation) over the course of the second cycle. We ran the simulation for two more cycles before introducing the perturbation, $\varepsilon=\SI{0.01}{\radian\per\second}$, to the system. As studied in \cite{Boyacioglu2021}, a regular cycle before the perturbation was necessary as the neural-encoded output has a delay, unlike the typical systems. The system was simulated one more cycle after the introduction of the perturbation, i.e., $t_1=\SI{40}{\milli\second}$.

\begin{figure}[t!]
\centering
\includegraphics[width=0.55\linewidth]{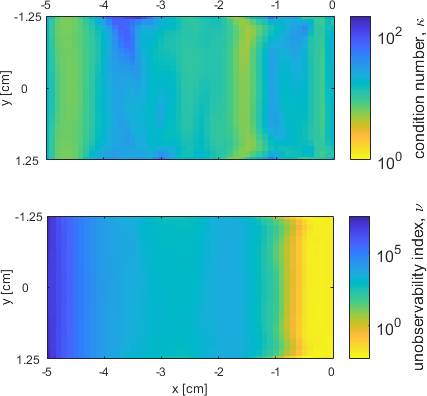}
\caption{The average distribution of two unobservability measures ($K=40$). Yellow regions are more observable than the dark blue ones.}\label{fig:sims_dist}
\end{figure}

We took the neural encoding parameters to be $a=\SI{5}{\milli\second}$, $b=\SI{4}{\milli\second}$, $\omega_\text{STA}=\SI{1000}{\radian\per\second}$, $c=10$, $d=0.5$, and $N=\SI{40}{\milli\second}$. The resultant distributions of the average unobservability index and the condition number stabilized in $K=40$ sets of simulations. In Fig. \ref{fig:sims_dist}, it can be seen that the unobservability index increases almost monotonically from the wing root ($x=0$) to the wingtip ($x=-5$). On the other hand, there are two separate regions where the eigenvalues are relatively more balanced. Finally, note that the asymmetric distributions about the $x$-axis are caused by the nonzero rotation rate.

\begin{figure}[hpbt!]
\centering
\includegraphics[width=0.6\linewidth]{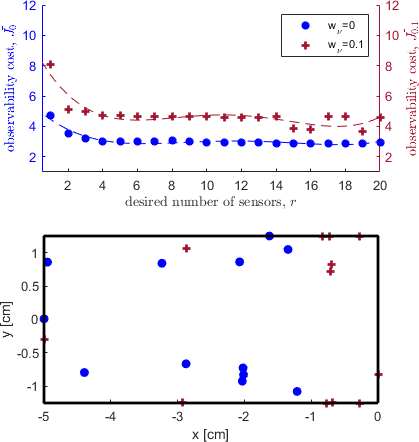}
\caption{(top) The change of the observability costs by the number of sensors with fourth-degree polynomial fitting, (bottom) The optimal neural-inspired sensor placement for $r=12$. The figures use the same legend.}\label{fig:sims_cost}
\end{figure}

We then ran the hybrid PSO algorithm for $r$ from one to $20$ to minimize the linear combination of $\kappa$ and $\nu$ with $w_\nu=0.1$ and to minimize solely the condition number, $w_\nu=0$. Since the search was performed in a continuous space, the strain value at a point was calculated as a weighted average of the strain values obtained for the neighboring FEA nodes. To avoid having sensors placed closer to each other than \SI{1}{\milli\metre}, we determined the penalty value as $\sigma=1\times10^5$.

The observability cost for increasing numbers of desired sensors is given in Fig. \ref{fig:sims_cost} (top). Both the condition number and its linear combination with the unobservability index converge, and adding a new sensor does not improve the observability significantly, which might be partially caused by the increasing dimension of the search.

Lastly, the optimal sensor placement for $r=12$ is shown in Fig. \ref{fig:sims_cost} (bottom). When the linear combination of the two measures with $w_\nu=0.1$ was utilized instead of a single unobservability measure, the sensors were slightly shifted to the left, and at least one sensor was placed at each edge. The use of pure condition number ($w_\nu=0$) resulted in having no sensors in the regions where the output energy is highest.

\section{CONCLUSIONS AND FUTURE WORK}\label{sec:last}

We formulated an optimal sensor placement methodology based on the stochastic empirical Gramian and illustrated it for two systems with process noise.  %a bioinspired system where the filter was embedded in the sensing mechanism. 
Since the problems are stochastic and non-convex, we performed Monte Carlo runs and used a metaheuristic optimization algorithm, PSO. We conclude that how the cost function is chosen matters, and a wise choice depends on the system structure and balances the output energy and the estimation condition.

%In this study, we have covered neither the estimation problem nor the neural decoding, which is widely studied in the field of computational neuroscience \cite{dayan2001theoretical}.
In light of the results from the UAV example, we plan to continue investigating how better observability measures result in better estimation performances. We will also study the convex relaxation of the optimization problem (\ref{eq:noncvxprob}) for various objective functions in the future. Higher moments of the probabilistic distribution (variance, skewness, and so on) might also be considered while posing the optimization problem, e.g., trust levels for sensor location candidates can be determined by the variance of the unobservability measures. Such an approach would be similar to the weighted least squares used in linear batch estimation \cite{Crassidis2011} or to the distributed least squares estimation in sensor networks \cite{mehran2010}.

A more realistic wing model can also be studied taking into account the actual wing profiles, nonuniform stiffness of the wing as in \cite{weber2023nonuniform}, and the structure of the strain-sensitive biological sensors. Finally, the neural spikes can be considered as the system output instead of the probability of firing, $P_\text{fire}$, and the possible sensor locations can be constrained to be on the vein of an insect wing.

Key points to be addressed in the next steps of the work are direct assessment of the improvements in filter performance from the optimal sensor placement,  exploration of the effects of different measures on the filter performance, and impacts of neural decoding methodology as expressed in the system measurement functions.

%These findings demonstrate a useful application of resolving the differences in observability between individual state variables, as opposed to only considering the observability of the full systems (which in this case is unobservable). Ideally, E-ISO can be applied to provide insights that might not otherwise be apparent from traditional observability tools.

%%%%%%%%%%%%%%%%%%%%%%%%%%%%%%%%%%%%%%%%%%%%%%%%%%%%%%%%%%%%%%%%%%%%%%%%%%%%%%%%

%%%%%%%%%%%%%%%%%%%%%%%%%%%%%%%%%%%%%%%%%%%%%%%%%%%%%%%%%%%%%%%%%%%%%%%%%%%%%%%%

%%%%%%%%%%%%%%%%%%%%%%%%%%%%%%%%%%%%%%%%%%%%%%%%%%%%%%%%%%%%%%%%%%%%%%%%%%%%%%%%
%\section*{APPENDIX}

\section*{Funding Sources}
This work was funded in part by Air Force Office of Scientific Research (AFOSR) grants FA9550-19-1-0386 and FA9550-14-1-0398.

\section*{Acknowledgement}

The authors would like to thank Alison I. Weber for the fruitful discussions on the flapping wing model and Natalie Brace for her constructive feedback on the manuscript draft.

\bibliographystyle{plainnat}
\bibliography{biblio}  %%% Uncomment this line and comment out the ``thebibliography'' section below to use the external .bib file (using bibtex) .

\end{document}